\documentclass{PoS}

\title{Measurement of the top quark properties in the production and decays of top anti-top pair events at CMS}

\ShortTitle{Top quark properties in  $t\bar{t}$ events at CMS}

\author{\speaker{Deniz POYRAZ}%
        \thanks{On behalf of the CMS Collaboration}\\
       Department of Physics and Astronomy, University of Ghent, Proeftuinstraat 86, B-9000 Ghent, Belgium\\
       E-mail: \email{deniz.poyraz@cern.ch}}

\abstract{Measurements of several top-quark properties obtained from  CMS data collected at various centre-of-mass energies are presented. The results include measurements of the top pair charge asymmetry,  $\textrm{t}\bar{\textrm{t}}$ spin correlation, CP violation and the cross section of $\textrm{t}\bar{\textrm{t}}$ events produced in association with a W or a Z boson. The results are compared with predictions from the standard model as well as new physics models.}

\FullConference{XXIV International Workshop on Deep-Inelastic Scattering and Related Subjects\\
		11-15 April, 2016\\
		DESY Hamburg, Germany}

\begin{document}

\section{Introduction}
The top quark is by far the heaviest  elementary particle observed with a mass of 173 GeV.  Due to its large mass, its coupling to the Higgs boson is very close to unity and this leads to the question whether the top quark has a significant role in electroweak symmetry breaking. Additionally its lifetime is shorter than the hadronization and the spin decorrelation time, ($1/\Gamma_t < 1/ \Lambda_{QCD}$) and ($1/\Gamma_t < m_t/\Lambda^2_{QCD}$), which  therefore  allows to study directly the bare quark properties.  The  top quark properties can be altered by  new physics, allowing to test the standard model (SM) and probe for beyond the standard model (BSM).
%beyond the standard models (BSM), therefore any deviation from the standard model (SM) prediction can be a sign of new physics.

%The top quark decays before it hadronizes,  ($1/\Gamma_t < 1/ \Lambda_{QCD}$), and  before its spin decorrelates,  ($1/\Gamma_t < m_t/\Lambda^2_{QCD}$). Therefore this allows to study directly the bare quark properties.   The top quark properties can be altered by new-physics therefore any deviation from the standard model prediction can be a sign of new physics.
%FIXME: add a sentence about top properties may be altered by some new physics models therefore it is a probe to study new physics so on...

\section{Charge Asymmetry}
The SM predicts a charge asymmetry in top pair production which is symmetric at the leading order (LO), and asymmetric at the next-to-leading order (NLO) due to QCD interference effects. Many new physics models predict a charge asymmetry at LO, therefore it is a powerful variable to constrain new physics.  The LHC has a charge-symmetric, proton-proton, initial state and the top quark pairs are produced in gluon-gluon fusion and quark-antiquark annihilation processes. The valance quarks carry a larger fraction of the proton momentum which leads top quarks to be produced  in forward-backward direction and anti-top quarks centrally. This asymmetry can be expressed as a function of the top quark pair  rapidities:

\begin{equation}
 = \frac{N(\Delta|y| > 0 ) - N(\Delta|y| < 0) }{N(\Delta|y| > 0) + N(\Delta|y| < 0)}
%A_C = \frac{N(|y_t| > |y_{\bar{t}}|) - N(|y_t| <  |y_{\bar{t}}|) }{N(|y_t| > |y_{\bar{t}}|) + N(|y_t| <  |y_{\bar{t}}|)}
\end{equation}

where $\Delta|y|~=~|y_t| -  |y_{\bar{t}}|$, or  by using the decay products of the top pairs as a function of the lepton pseudo-rapidities in dilepton final state:

\begin{equation}
A^{lep}_C  = \frac{N(\Delta|\eta| > 0 ) - N(\Delta|\eta| < 0) }{N(\Delta|\eta| > 0) + N(\Delta|\eta| < 0)}
%A^{\ell\ell}_C = \frac{N(|y_t| > |y_{\bar{t}}|) - N(|y_t| <  |y_{\bar{t}}|) }{N(|y_t| > |y_{\bar{t}}|) + N(|y_t| <  |y_{\bar{t}}|)}
\end{equation}

where $\Delta|\eta|~=~|\eta_{\ell^+}| -  |\eta_{\ell^-}|$.  Theoretical predictions for these values are $\sim~1\%$ in the SM.\\

\begin{figure}
\centering
\includegraphics[width=.45\textwidth]{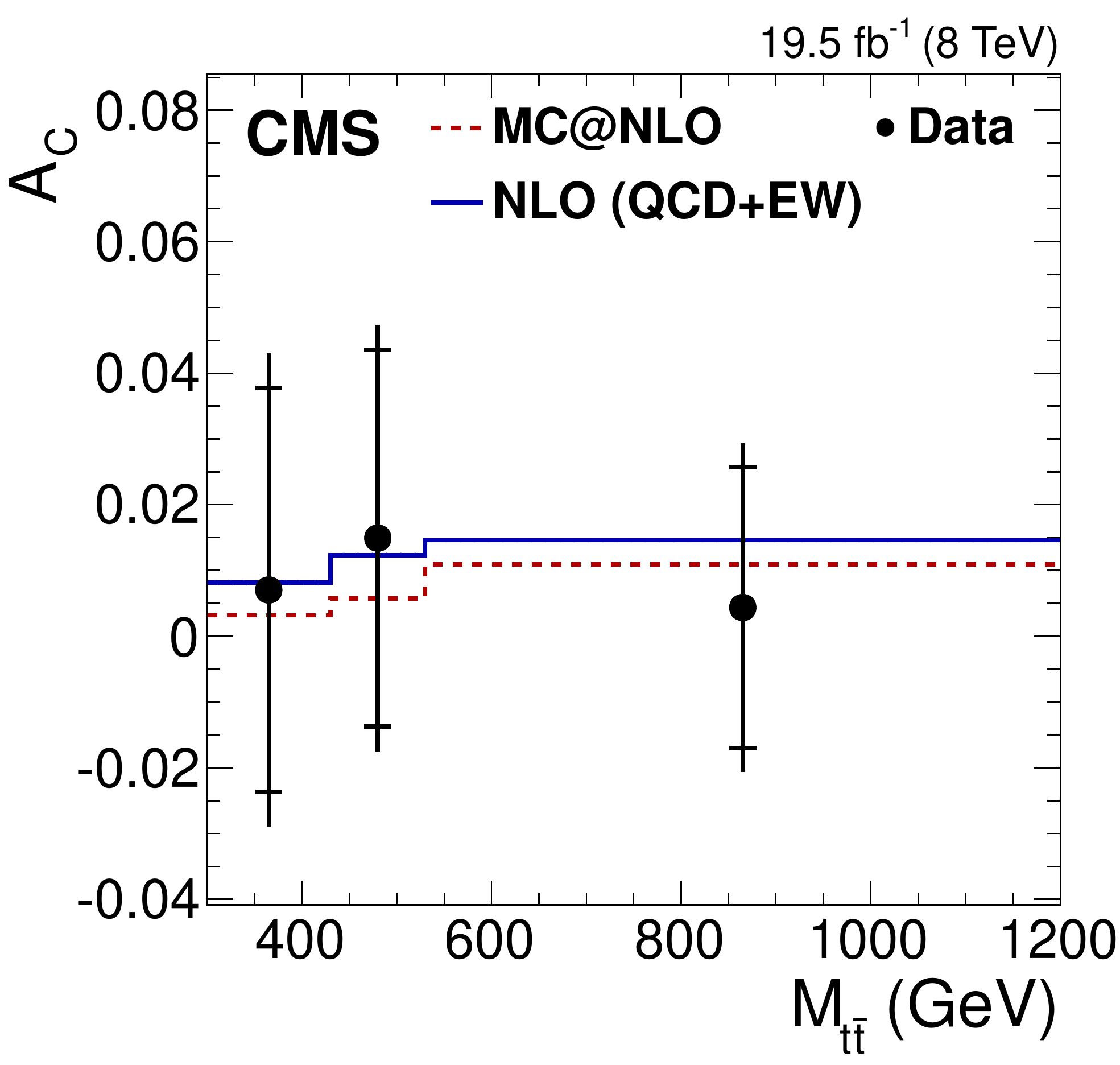}
 \includegraphics[width=.45\textwidth]{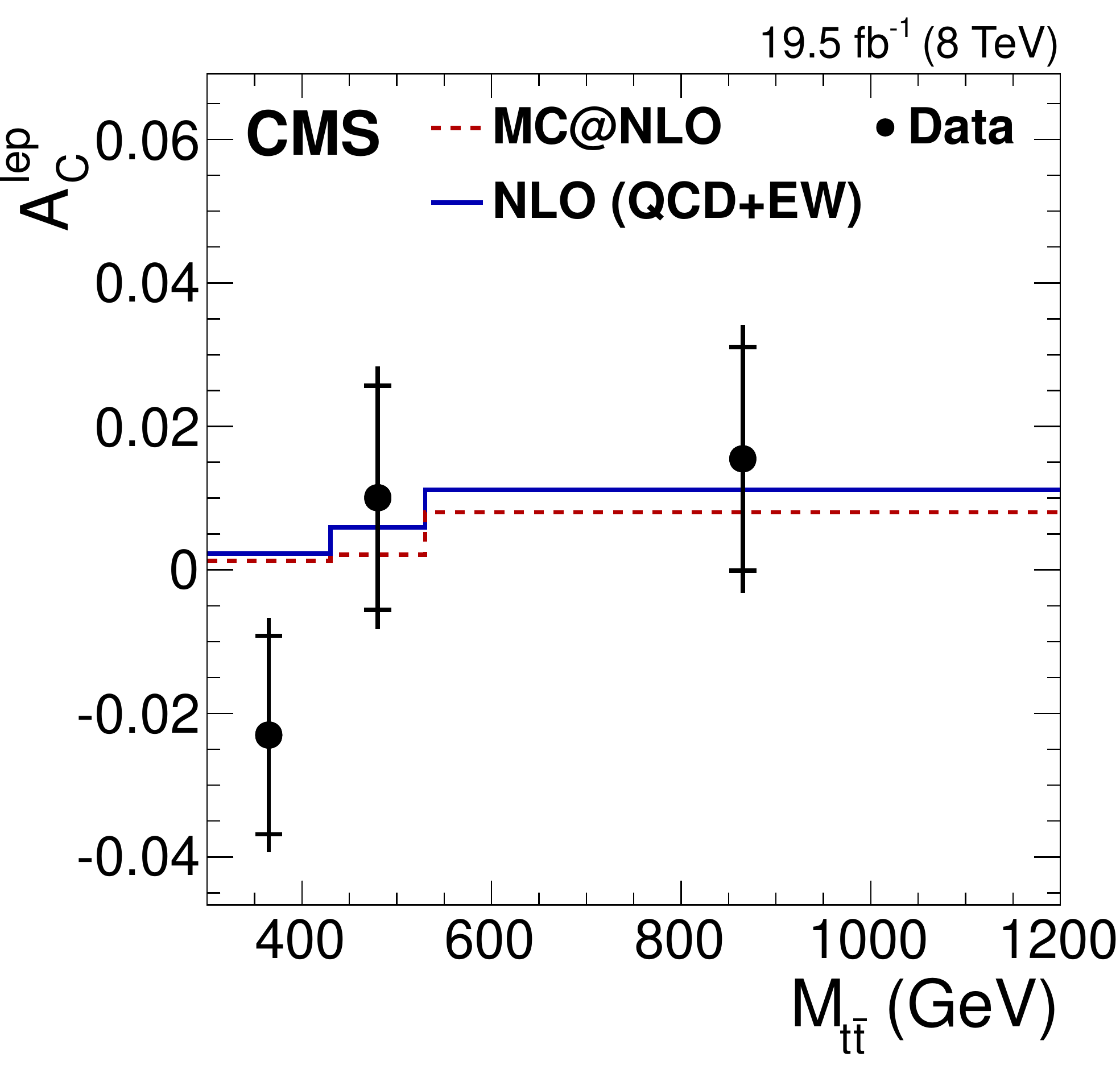}  
\caption{Dependence of the  $\textrm{t}\bar{\textrm{t}}$   and leptonic charge asymmetries $A_C$ (left) and $A^{lep}_C$ 
obtained from the unfolded distributions in data (points) on $M_{\textrm{t}\bar{\textrm{t}}}$ \cite{Khachatryan:2016ysn} . Parton-level predictions from the  MC@NLO simulation and calculations at NLO (QCD+EW)~\cite{Bernreuther:2012sx} are shown by dashed and solid histograms, respectively.}
\label{fig:CA}
\end{figure}

The CMS~\cite{cms} collaboration  measured the  $\textrm{t}\bar{\textrm{t}}$  charge asymmetry with different methods and in different final states at 8 TeV with 19.7 fb$^{-1}$ of data.  The first measurement is done in  single-lepton final state events, with an unfolding procedure applied for detector effects, inclusively and  differentially as a function of the $\textrm{t}\bar{\textrm{t}}$ system rapidity $|y_{\textrm{t}\bar{\textrm{t}}}|$, transverse momentum $p^{\textrm{t}\bar{\textrm{t}}}_{T}$ and invariant mass $m_{\textrm{t}\bar{\textrm{t}}}$ ~\cite{Khachatryan:2015oga}. The results  are $A_C = [-0.35\pm0.72(\textrm{stat).}\pm0.31(\textrm{sys.})]\%$ and $A_C = [0.10\pm0.68(\textrm{stat.})\pm0.37(\textrm{sys.})]\%$ for the inclusive full space and fiducial space measurements respectively. The differential measurements are in agreement with the SM and the results are compared with predictions from an effective field theory  that involves the effective axial-vector coupling of the gluons~\cite{Gabrielli:2011zw, Gabrielli:2011jf}. The measurement excludes new physics scales below 1.5 TeV at the $95\%$ confidence level. Another inclusive measurement is done in the same final state  with a template technique ~\cite{Khachatryan:2015mna} based on a parametrization of the SM. The result  $A_C = [0.33\pm0.26(\textrm{stat.})\pm0.33(\textrm{sys.})]\%$  is the most precise measurement up to date.  The charge asymmetry measurement is also done in the dilepton final state using the unfolding method \cite{Khachatryan:2016ysn} . As opposed to the single-lepton analyses, the dilepton analyses also includes an asymmetry calculation based on the lepton pseudo-rapidities
%The difference of this analysis with respect to other analyses is the asymmetry is calculated using the lepton pseudo-rapidities as well as the $\textrm{t}\bar{\textrm{t}}$ system rapidity. 
The results  are $A^{lep}_C = [0.3\pm0.6(\textrm{stat.}\pm0.3(\textrm{sys.})]\%$ and $A_C = [1.1\pm1.1(\textrm{stat.})\pm0.7(\textrm{sys.})]\%$ for the inclusive full space measurements, the results can be seen in Fig.~\ref{fig:CA}. All the measurements are in agreement with the SM. 

\section{Spin Correlations}
The top quarks decay before their spins decorrelate,  therefore their spin correlation is propagated to the  $\textrm{t}\bar{\textrm{t}}$ decay products. The SM predicts the $\textrm{t}\bar{\textrm{t}}$  spins to be correlated and the spin correlation strength can be defined as the asymmetry between the number of  $\textrm{t}\bar{\textrm{t}}$ pairs with aligned and anti-aligned pairs.

%\begin{equation}
%A  = \frac{ (N_{\uparrow\uparrow} + N_{\downarrow\downarrow}) - (N_{\uparrow\downarrow} + N_{\downarrow\uparrow}) } { (N_{\uparrow\uparrow} + N_{\downarrow\downarrow})  + (N_{\uparrow\downarrow} + N_{\downarrow\uparrow})	}
%\end{equation}

The first analysis from CMS at 8 TeV with 19.7 fb$^{-1}$ of data is in muon+jets final state, and the measurement is done by using a matrix element method~\cite{Khachatryan:2015tzo}. 
The discriminating variable used for the event likelihood  fit is the ratio of the SM (correlated) and uncorrelated  $\textrm{t}\bar{\textrm{t}}$  pairs. The fraction of the SM $\textrm{t}\bar{\textrm{t}}$  pairs is measured to be $f^{SM} = 0.72\pm 0.09(\textrm{stat.}) ^{+0.15}_{-0.13}(\textrm{sys.})$. The second analysis from CMS is, also at 8 TeV with 19.7 fb$^{-1}$ of data,  performed in the dilepton final state using the angular distributions and  asymmetry variables of the leptons  unfolded to the parton level and as a function of the  $\textrm{t}\bar{\textrm{t}}$ system variables  $m_{\textrm{t}\bar{\textrm{t}}}$,  $|y_{\textrm{t}\bar{\textrm{t}}}|$ , and  $p^{\textrm{t}\bar{\textrm{t}}}_{T}$ \cite{Khachatryan:2016xws}. The fraction of the SM $\textrm{t}\bar{\textrm{t}}$  pairs is measured to be $f^{SM} = 1.12^{+0.12}_{-0.15}$. 

\section{CP Violation}
\begin{figure}
\centering
\includegraphics[width=.45\textwidth]{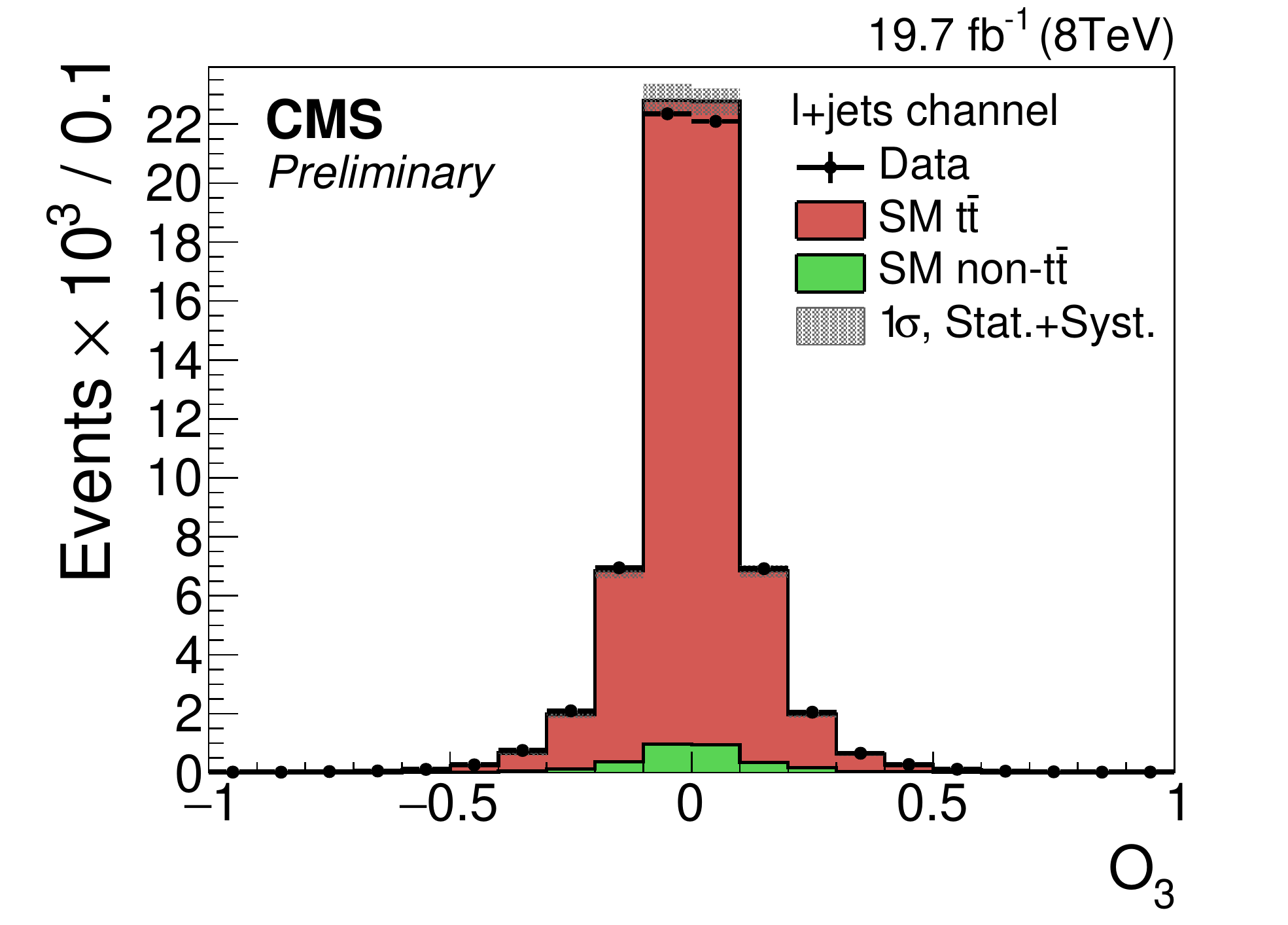}
 \includegraphics[width=.45\textwidth]{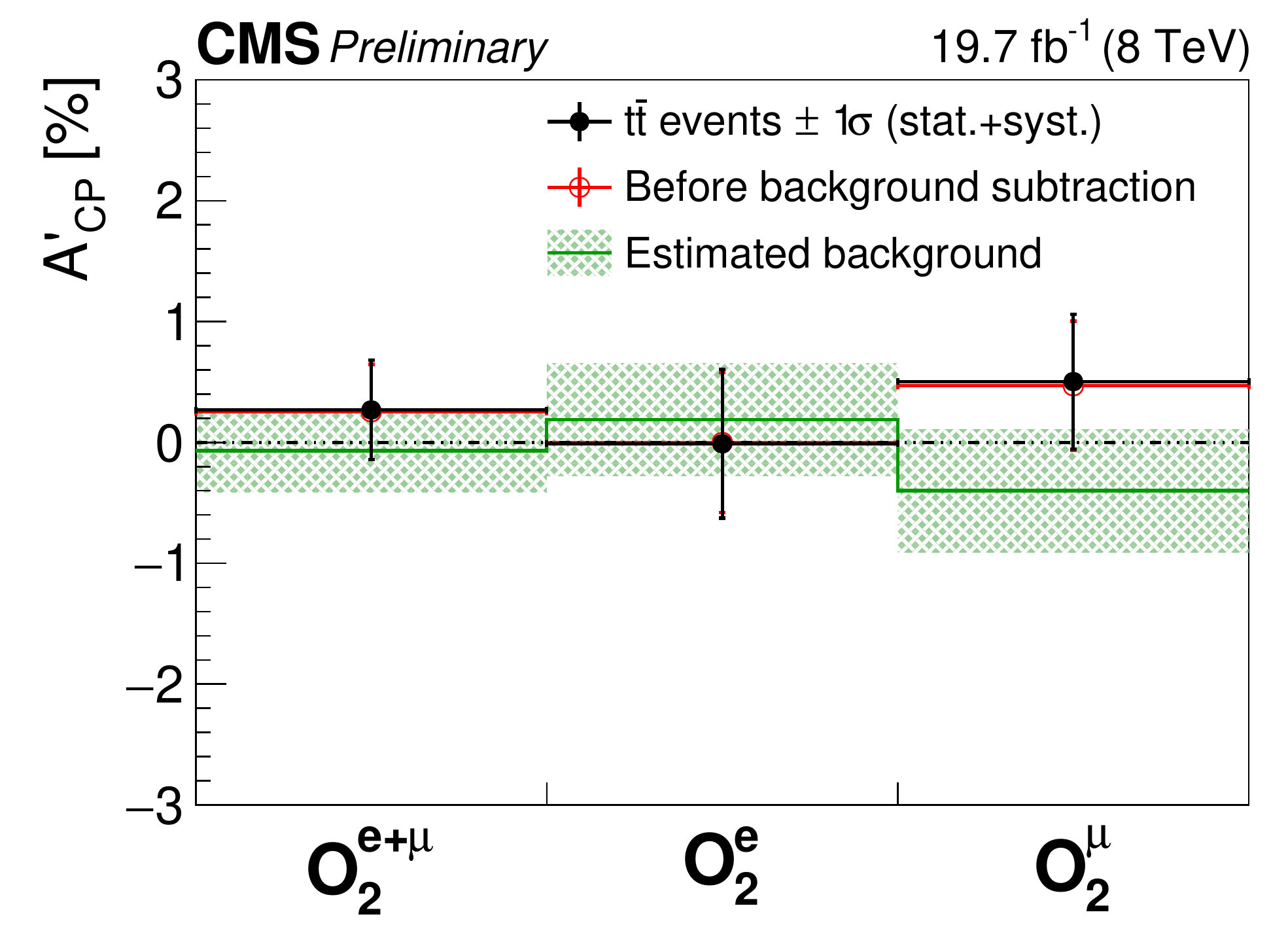}  
\caption{Distribution of the observables of data and simulation (SM) in the signal region in combined channels before the background subtraction~\cite{CMS:2016tnr}. Each observable is given in the units of $m_t^3$ (left). Summary of the $A_{CP} $  measurements performed for one observable in all the channels (right). }
\label{fig:CP}
\end{figure}

In the SM, charge conjugate and parity symmetry (CP) violation in the $\textrm{t}\bar{\textrm{t}}$  production pairs is predicted to be very small. The CMS collaboration has measured the CP violation in $\textrm{t}\bar{\textrm{t}}$  production and decay  at 8 TeV with 19.7 fb$^{-1}$ of data~\cite{CMS:2016tnr}.  The measurement is done by exploiting the T-odd triple product observables, which are odd under CP transformation. The CP violation can be assessed by measuring a non-zero value of the asymmetry:
\begin{equation}
A_{CP} (O_i) = \frac{N_{events}(O_i > 0 ) -  N_{events}(O_i < 0 )  } { N_{events}(O_i > 0 ) +  N_{events}(O_i < 0 ) }
\end{equation}
where $O_i$ are the observables as a function of the four-momenta of the decay products. The asymmetry is calculated as a function of these observables and no evidence for CP-violation is observed. The distribution of an observable and $A_{CP}$ as a function of one observable can be seen in Fig.~\ref{fig:CP}.
 
\section{$\textrm{t}\bar{\textrm{t}}$  +V production cross sections}
\begin{figure}
\centering
\includegraphics[width=.5\textwidth]{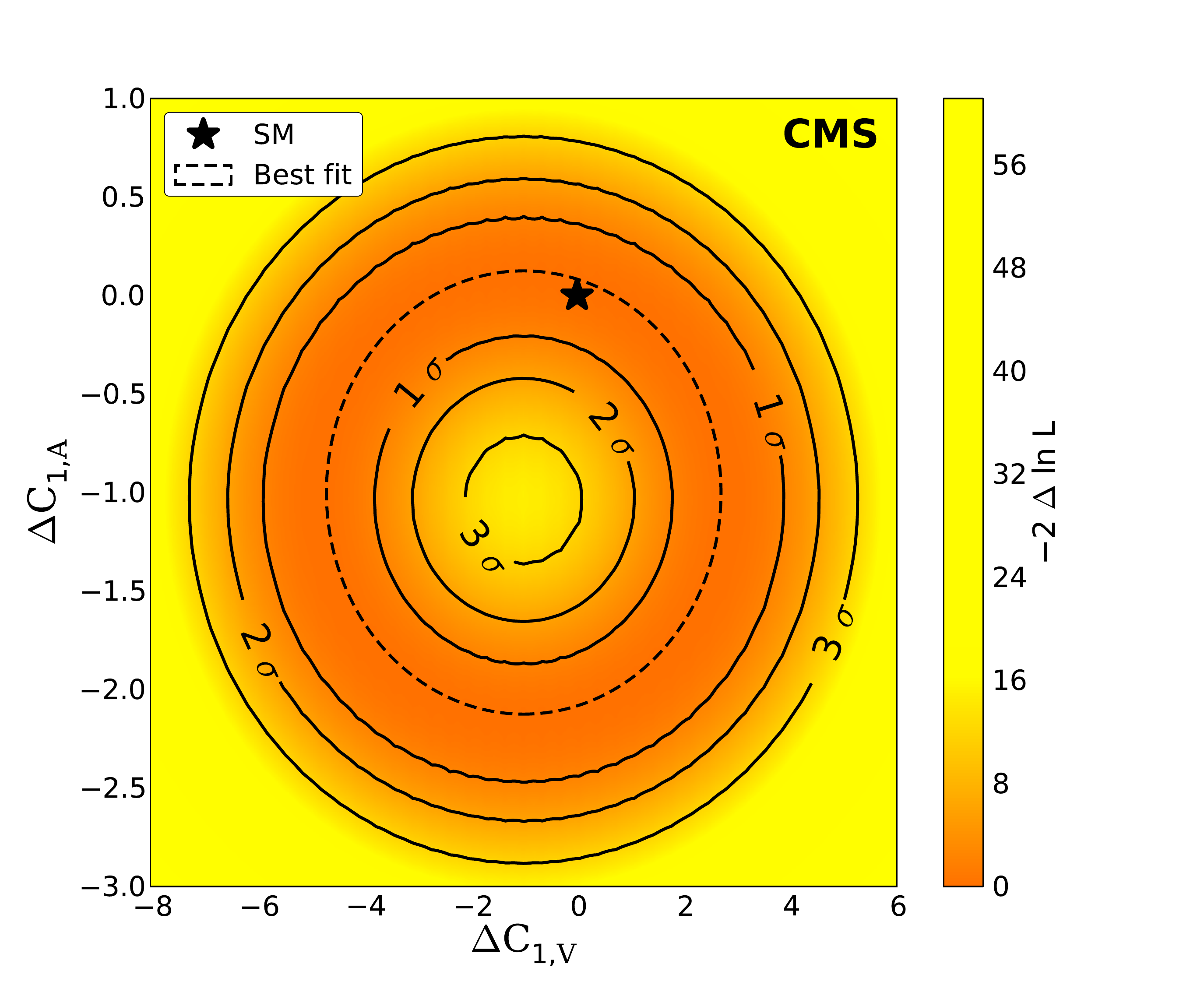}
 \includegraphics[width=.425\textwidth]{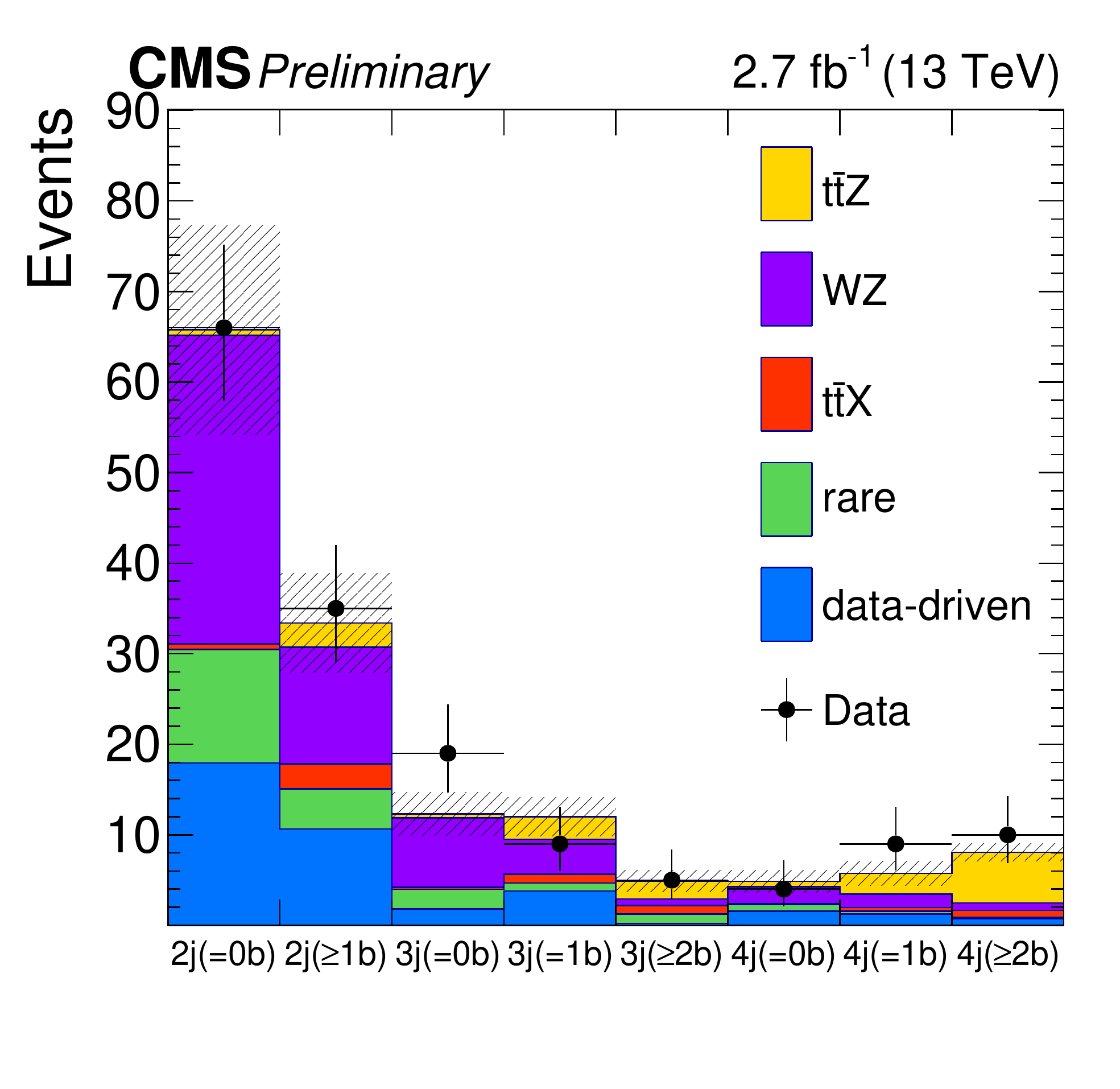}  
\caption{Difference between the profile likelihood and the best fit profile likelihood functions for the relative vector and axial components of the tZ coupling, contours corresponding to the best fit and the 1, 2, and 3 standard deviation CLs are shown in lines (left) ~\cite{Khachatryan:2015sha}. Number of events in the $\textrm{t}\bar{\textrm{t}}$Z three-lepton signal regions (right)~\cite{CMS:2016ium}. }
\label{fig:ttV}
\end{figure}

The  $\textrm{t}\bar{\textrm{t}}$V measurement by CMS at 8 TeV with 19.7 fb$^{-1}$ of data is done by using  event reconstruction discriminators with a multivariate analysis  in  all possible leptonic final states~\cite{Khachatryan:2015sha}. The first observation of  $\textrm{t}\bar{\textrm{t}}Z$ and measurement of its cross section with 6.4 standard deviations is reported to be $\sigma_{\textrm{t}\bar{\textrm{t}}Z} = 242^{+65}_{-55}~\textrm{fb}$. The $\textrm{t}\bar{\textrm{t}}W$ cross section is measured with 4.8 standard deviations, $\sigma_{\textrm{t}\bar{\textrm{t}}W} = 382^{+117}_{-102}~\textrm{fb}$. Additionally constraints on the axial and vector components of the tZ coupling and on dimension-six operators were put. The other analysis presented by the CMS collaboration is measurement of  $\textrm{t}\bar{\textrm{t}}$Z cross section at 13 TeV with 2.7 fb$^{-1}$ of data~\cite{CMS:2016ium}. The measurement is done in three- and four-lepton final states with a binned likelihood fit to extract cross section and measured $\sigma_{\textrm{t}\bar{\textrm{t}}Z} = 1065^{+352}_{-313}(\textrm{stat.})^{+168}_{-142}(\textrm{sys.})~\textrm{fb}$ with 3.6 standard deviations. The limits on the tZ coupling from the first analysis and the number of selected events in three lepton analysis from the second analysis can be seen in Fig.\ref{fig:ttV}.

\end{document}